\documentclass[aps,prl,twocolumn]{revtex4-2}

\usepackage{graphicx}
\usepackage{dcolumn}
\usepackage{bm}
\usepackage{amsmath}
\usepackage{amssymb}
\usepackage{url}

\usepackage{color}
\usepackage[colorlinks,bookmarks=false,citecolor=darkblue,linkcolor=red,urlcolor=blue]{hyperref}

\definecolor{darkred}{rgb}{0.7,0.0,0.0}
\definecolor{darkblue}{rgb}{0,0.02,0.45}

\graphicspath{{figures/}}

\begin{document}

\title{Anomalous vortex dynamics in spin-triplet superconductor UTe$_2$}

\author{Y. Tokiwa*$^{1}$}
\author{H. Sakai$^{1}$}
\author{S. Kambe$^{1}$}
\author{P. Opletal$^{1}$}
\author{E. Yamamoto$^{1}$}
\author{M. Kimata$^{2}$}
\author{S. Awaji$^{2}$}
\author{T. Sasaki$^{2}$}
\author{Y. Yanase$^{3}$}
\author{Y. Haga$^{1}$}
\author{Y. Tokunaga$^{1}$}

\affiliation{$^1$Advanced Science Research Center, Japan Atomic Energy Agency, Japan.}
\affiliation{$^2$Institute for Materials Research, Tohoku University, Sendai, Miyagi, Japan}
\affiliation{$^3$Department of Physics, Kyoto University, Kyoto 606-8502, Japan}

\affiliation{{\rm * ytokiwa@gmail.com}}
\date{\today}

\begin{abstract}
The vortex dynamics in the spin-triplet superconductor, UTe$_2$, are studied by measuring the DC electrical resistivity with currents along the $a$-axis under magnetic fields along the $b$-axis. Surprisingly, we have discovered an island region of low critical current deep inside the superconducting (SC) state, well below the SC upper critical field, attributed to a weakening of vortex pinning. Notably, this region coincides with the recently proposed intermediate-field SC state. We discuss the possibility of nonsingular vortices in the intermediate state, where SC order parameter does not vanish entirely in the vortex cores due to the mixing of multiple SC components.
\end{abstract}

\maketitle

Unconventional superconductivity often leads to an unconventional vortex states. In conventional cases, the thermodynamic superconducting (SC) transition and the formation of vortex lattice occur simultaneously at $T_c$. However, in unconventional superconductors,  these could appear sequentially due to their anomalously large $T_c$ compared to their condensation energy per coherent volume, leading to the formation of a vortex liquid state in a region below $T_c$~\cite{Blatter}. Since the lattice constant of vortex continuously decreases with increasing magnetic field, a vortex lattice melts above a magnetic field close to $B_{\rm c2}$, when the amplitude of vortex lattice fluctuations becomes comparable to the lattice constant~\cite{Blatter}.

Among heavy fermion compounds, URu$_2$Si$_2$ and UCoGe exhibit unusually wide regions of vortex liquid states~\cite{Okazaki2008,Wu2018}.  For URu$_2$Si$_2$, the capability of growing ultra-clean crystals clearly contributed to the observation of liquid state~\cite{Okazaki2008}. The situation is more complex for UCoGe, since the compound exhibits spin-triplet SC and the vortex liquid state is accompanied by a field-reinforcement of superconductivity for the field applied along a magnetically hard-axis~\cite{Wu2018}.
There, the weakening of the pinning force due to the formation of nonsingular fractional vortices has been discussed~\cite{Wu2018}. Such fractional vortices may host Majorana fermions, which obey non-Abelian statistics~\cite{Read2000,Ivanov2001}. These studies stimulate our interest in investigating the vortex dynamics in the spin-triplet superconductor UTe$_2$~\cite{Ran2019,Aoki2019,Knebel2019,Aoki2020a,Kinjo2022,Aoki2022a,Rosuel2022,Fujibayashi2022,Braithwaite2019,Aoki2020a,Lin2020,Jiao2020,Kittaka2020}, which was recently shown to have multiple SC states under magnetic fields along the magnetic hard $b$-axis~\cite{Braithwaite2019,Aoki2020a,Rosuel2022,Kinjo2022,Sakai2022arxiv}.

For UTe$_2$, significant efforts have been made to improve crystal quality~\cite{Haga2022,Rosa2022a,Sakai2022}. The recent progress of growing clean crystals has led to the observation of quantum oscillations~\cite{Sakai2022,Aoki2022c,Eaton2023,Broyles2023,Aoki2023} and renewed the SC phase diagram with higher $B_{c2}$~\cite{tokiwa2022arxiv, Sakai2022arxiv}.
The reinforcement of superconductivity occurs for the field along the $b$-axis, where the applied field initially suppresses superconductivity, but enhances $T_c$ above $B^\star\sim$15 T, causing a minimum of $T_c$($B$) at $B^\star$~\cite{Ran2019,Knebel2019,Rosuel2022}. The recent studies have revealed the presence of a phase boundary within the SC state for $B\|b$, which divides the superconductivity into low and high-field SC states~\cite{Rosuel2022,Kinjo2022}. Most recently, another phase boundary has been found in the low-field SC state around $B^\star$, implying the presence of in-total three SC states with the intermediate SC (IFSC) state, which may be characterized by mixture of multiple SC order parameters from the low- and high-field SC states (LFSC and HFSC states)~\cite{Ishizuka2019,Sakai2022arxiv}.

In this Letter, using an ultra-clean single crystal  grown by the molten-salt flux method~\cite{Sakai2022}, we study the vortex dynamics in the multiple SC states of UTe$_2$.
Remarkably, our direct current (DC)-electrical resistivity measurements with different currents reveal a sudden weakening of the pinning force in the IFSC state for $B\|b$~\cite{Sakai2022arxiv}. This may be caused by the formation of nonsingular vortices in the IFSC state, where the SC order parameter does not vanish entirely in vortex cores due to the mixing of multiple SC components.

DC-resistivity measurements were performed using the four-probe method and a $^3$He cryostat with an SC magnet in the High-Field Laboratory for Superconducting Materials at the Institute for Materials Research at Tohoku University. The sample was placed inside $^3$He liquid during the measurements. The rod-shaped sample used for the measurements has a cross-sectional area of 0.29$\times$0.26$=$0.075\,mm$^2$ and a length of 2.2\,mm along the $a$-axis between the voltage contacts. Magnetic field was applied along the $b$-axis. The sample,  grown by molten-salt flux method, has a SC transition temperature of $T_c$=2.1\,K\,\cite{Sakai2022}. The Joule heating effect was examined by applying current in the normal conducting state at a temperature of 1.46\,K and in magnetic field of 20\,T. The voltage shows no deviation from the linear dependence of the electrical current ($I$) within the experimental accuracy up to 60\,mA. Above this current, it deviates with an upward curvature due to heating. We can neglect the Joule heating effect because the upper bound of the ohmic region, 60\,mA, is 10 times larger than the typical current, 6\,mA and larger than the maximum current of 45 mA for measurements of $E-J$ characteristics.

\begin{figure}
\includegraphics[width=\linewidth,keepaspectratio]{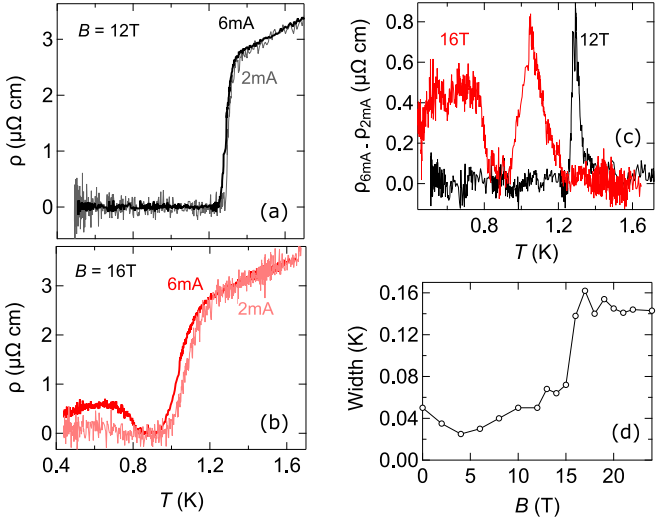}%
\caption{\label{t_dep} DC-resistivity of UTe$_2$ measured with different currents, 2\,mA (grey, pink) and 6\,mA (black, red) along $a$-axis under magnetic fields of 12\,T (a) and 16\,T (b) along $b$-axis. (c) The difference between the resistivity with the two currents, $\rho_{\rm 6mA}-\rho_{\rm 2mA}$, at $B$=12 and 16 T.  (d) Magnetic field dependence of the full width of half maximum for the peak observed in $\rho_{\rm 6mA}-\rho_{\rm 2mA}$ at the normal conducting-superconducting transition temperature. }
\end{figure}

Figures \ref{t_dep}(a,b) show the DC-resistivity with two different currents, 2 and 6\,mA along the $a$-axis under magnetic fields of 12 and 16\,T along the $b$-axis. While the SC transition is sharp at $B$=12\,T, it broadens significantly at 16\,T. The data with 6\,mA at 16\,T shows zero resistance below 1\,K and remarkably recovers nonzero resistance below 0.8\,K. The difference between $\rho$ measured by the two different currents, $\rho_{\rm 6mA}-\rho_{\rm 2mA}$, is plotted in Fig. \ref{t_dep}(c). A peak at the normal conducting (NC)-SC transition temperature, $T_c$, is clearly broader for $B$=16\,T than for $B$=12 T. The full width at half maximum of the peak at $T_c$ is plotted in Fig.~\ref{t_dep}(d). The position of step-wise increase, $B$=15\,T, coincides with the multi-critical point of different SC states, $B^\star$~\cite{Kinjo2022,Rosuel2022,Sakai2022arxiv}. This result is consistent with the reported broadening of the transition for the high-field reinforced SC state~\cite{Rosuel2022,Sakai2022arxiv}.

\begin{figure}
\includegraphics[width=\linewidth,keepaspectratio]{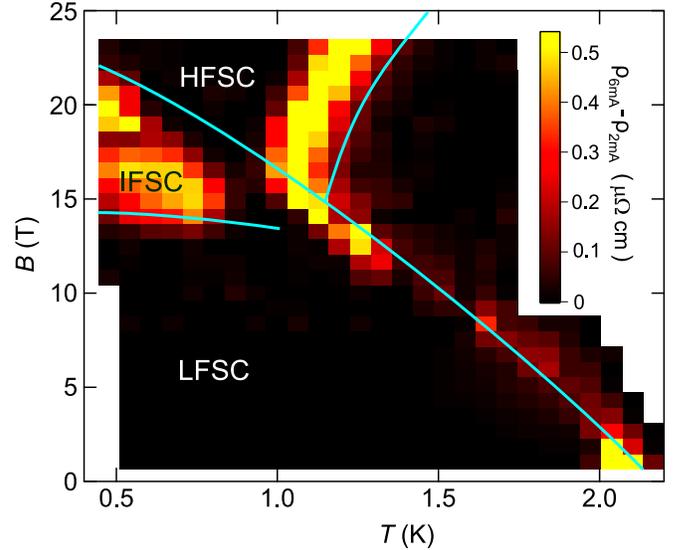}%
\caption{\label{contour} Color contour plot of the difference $\rho_{\rm 6mA}-\rho_{\rm 2mA}$. The plot is constructed from the temperature-dependent DC-resistivity $\rho(T)$ measurements at various magnetic fields. $\rho(T)$ at 2 and 6 mA are taken at magnetic fields from 0 to 12 T with an interval of $\Delta$$B$=2 T, from 13 to 22 T with  $\Delta$$B$=1 T and taken finally at 24 T. The solid blue lines are the phase boundaries determined by the recent study~\cite{Sakai2022}. "LFSC", "IFSC" and "HFSC" denote the low-field, intermediate-field and high-field superconducting states, respectively~\cite{Sakai2022}.}
\end{figure}

From resistivity measurements at various magnetic fields, we constructed the color contour plot of $\rho_{\rm 6mA}-\rho_{\rm 2mA}$ as shown in Fig.~\ref{contour}. It reproduces the peculiar NC-SC phase boundary with the reinforcement of superconductivity above $B^{\star}\sim$15\,T. Our $T_c(B)$ agrees well with the reported one for low fields below $B^\star$, whereas it is slightly shifted to lower temperatures by $\sim$0.2\,K at high fields above $B^{\star}$. The shift only for HFSC state is explained by a sample miss-alignment of $\sim$2 degree from $B\parallel b$-axis, because $T_c$ for the HFSC state is very sensitive to the field angle, whereas that for the LFSC state is much less sensitive~\cite{Sakai2022arxiv}.

\begin{figure}
\includegraphics[width=\linewidth,keepaspectratio]{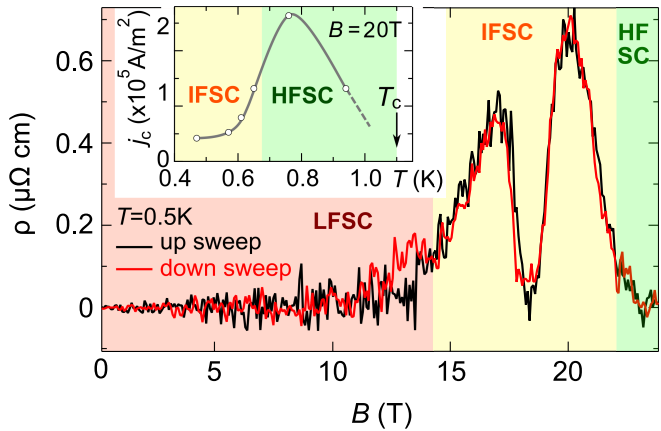}%
\caption{\label{B_dep} Magnetic field dependence of the resistivity of UTe$_2$ at a temperature of 0.5 K for the field along the $b$-axis with a current of 6 mA along the $a$-axis. The black and red curves are the data taken by sweeping magnetic field up and down, respectively.  The red, yellow and green colored regions indicated by "LFSC", "IFSC" and "HFSC" are the low-field, intermediate-field and high-field superconducting states, respectively. Inset shows the critical current density, $j_c$, at $B$=20\,T which is needed to induce finite resistivity. ``$T_c$" denotes the normal conducting (NC)-superconducting phase transition determined by zero resistance with a small current of 0.5 mA.}
\end{figure}

The peculiar behavior of $\rho$($T$) with 6\,mA at $B$=16\,T indicates that the vortices form a solid below $T_c$ and start to flow below 0.8\,K. For 2\,mA vortices remain solid down to 0.45 K. The vortex-flow state for 6 mA corresponds to the finite $\rho_{\rm 6mA}-\rho_{\rm 2mA}$ region deep inside the SC state, which spans magnetic fields of 14-21\,T and temperatures below 0.8\,K. Here, it should be mentioned that the color coding in Fig.~\ref{contour} does not represent the vortex-flow resistivity, $\rho_f$=$dE/dj$, where $E$ and $j$ are the electric field and current density, respectively. It rather highlights the region of low critical current because the light color indicates the region of critical current lower than 6 mA. More detailed investigations on the vortex-flow resistivity are left for future studies. As shown in Fig~\ref{t_dep}(c), $\rho_{\rm 6mA}-\rho_{\rm 2mA}$ at 16\,T exhibits the two vortex-flow states right below $T_c$ and below 0.8 K, separated by a solid state around 0.9 K. Such a separation is always observed in any magnetic fields of our $\rho(T)$ measurements between 12 and 22\,T, with an interval of 1 T. Because thermal fluctuations increase with temperature, fluctuation-induced vortex-flow would not disappear with increasing temperature but rather persists up to $T_c$. Therefore, the formation of island region of low critical current deep inside the SC state is ascribed to weakening of vortex pinning.  Reflecting the separation, the critical current density $j_c$, needed to induce finite resistivity, exhibits an unusual maximum as a function of temperature and decreases on entering the IFSC state (the inset of Fig.~\ref{B_dep}).

Notably, there is a narrow field range around 18 T of $\rho_{\rm 6mA}-\rho_{\rm 2mA}=0$ in the middle of the low $j_c$ region. This separates the low $j_c$ region into lower and higher field ones, spanning 14-17 and 19-21 T, respectively. This is not an experimental error but rather reproducible. The magnetic field dependence of $\rho_{\rm 6mA}$ clearly indicates vortex solidification with $\rho_{\rm 6mA}=$0 at 18\,T between the two flow states with nonzero $\rho_{\rm 6mA}$ (Fig.~\ref{B_dep}). There is no clear hysteresis between the data for sweeping magnetic field up and down. The field range of solidification is independent of temperature. Therefore, it might be caused by a kind of matching effect of vortices. 

\begin{figure}
\includegraphics[width=0.9\linewidth,keepaspectratio]{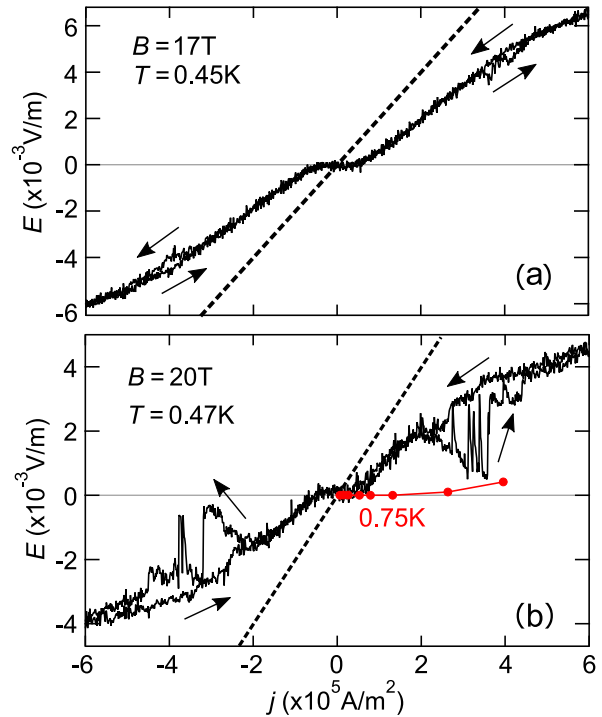}%
\caption{\label{jc} Electric field-current ($E-J$) characteristics deep inside SC state at $B$=20 and 17 T. Dotted lines represent the resistivity of NC state, which is determined by extrapolating NC resistivity with $T^2$-dependence to the measurement temperatures for $E-J$ characteristics. Data at $T$=0.75 K for $B$=20 T is plotted for comparison (red). The maximum current density corresponds to a current of 45 mA.}
\end{figure}

Figures~\ref{jc}(c,d) show the $E-j$ characteristics in the two regions of the low $j_c$. They indicate a narrow $E=0$ region of the vortex solid state for $|j|<4\times$10$^4$ A/m$^2$ (3 mA). On the other hand, at an elevated temperature of 0.75 K for $B$=20 T, the critical current density 27$\times$10$^4$ A/m$^2$ to induce non-zero $E$ is much larger. Interestingly, both data sets at 17 and 20 T display unusual current-density dependence. At 20 T, $E$ increases with $j$ at low $j$ region, but turns to decrease above $j$=2$\times$10$^5$ A/m$^2$ and almost reaches $E=0$ (solidification). Furthermore, we note that $E$ shows repeated sharp changes between the low and high values in the $j$ range. This indicates that less-movable vortices suddenly flow at random $j$ values, released from vortex pinning. This behavior is reproducible in the reversed current direction ($j<0$) and occurs only for sweeping up the magnitude of the current density $|j|$, resulting in the hysteresis. At 17 T, the same but much weakened behavior is observed at a slightly elevated $|j|$ range. Additionally, the slope becomes smaller in the same $|j|$ range. The origin of the stronger pinning force in this $|j|$ range calls for future studies.

We discuss the possible origins of the weakening of the pinning force deep inside the SC state. The vortex-flow region shown in Fig. \ref{contour} coincides well with the recently proposed IFSC state, which has a lower-field phase boundary around 14\,T\cite{Sakai2022arxiv}. Therefore, the weaker pinning force arises from the properties of the IFSC state ~\cite{Ishizuka2019,Sakai2022arxiv}.

With the orthorhombic crystal structure of UTe$_2$,  the SC states are classified into $A_u^{\|b}$ or $B_u^{\|b}$ irreducible representations when the magnetic field is applied along the $b$ axis \cite{Ishizuka2019}. Using  the classification of the odd-parity SC order parameters for the point group $D_{2h}$ at zero field,
the $A_u^{\|b}$ and $B_u^{\|b}$ states are represented as $A_u^{\|b}=A_u+ B_{2u}$ and $B_u^{\|b}=B_{1u}+ B_{3u}$, respectively.
For the SC gap structure, the former has symmetry protected point nodes on the $k_y$ axes, whereas the latter has a line node on the $k_y=0$ plane. Although it has not been settled which of the two representations corresponds to the LFSC or the HFSC states, the change in the SC gap structure would result in a change in the pinning force by modifying the interaction between vortices and impurities/defects.
The anisotropy of the SC gap influences both the vortex core and lattice structure, as the latter discussed for the multiple SC states of UPt$_3$~\cite{Ichioka1999,Huxley2000}.

Furthermore,  if  the IFSC state could be characterized by a mixing of the multiple SC components from the LFSC and HFSC states, i.e. $A_u^{\|b}+B_u^{\|b}$,
as a consequence that, the formation of nonsingular, coreless vortices may be expected~\cite{Mermin1976,Anderson1977,Salomaa1987,Sigrist1991,Ivanov2001,Autti2016}. Such coreless (fractional) vortices, have been theoretically discussed, for example, in a field-induced chiral state of Sr$_2$RuO$_2$~\cite{Takamatsu2013}, where the corresponding chiral state is characterized by degenerate pairing states with multiple (=4) components, and the fractional vortex lattice is suggested to be stabilized by the spin-orbit coupling.
In general, vortices are pinned at impurities/defects, because the energy cost is minimized when the NC vortex cores are positioned at the impurities/defects, where the superconductivity is weaker or absent.
In a nonsingular vortex state with multiple SC components, different components form vortex lattices separately, with the real-space positions of vortex cores shifted relative to each other~\cite{Chung2009,Takamatsu2013}. This results in the non-zero SC order parameter everywhere, reducing the pinning force. In this case, several possibilities would explain the observed pinning around $B=18$T. One possibility is a matching between the lattice constants of the two vortex lattices, which may have different field dependencies. Another possibility is a first-order vortex-lattice phase transition, unique in such a mixed SC state. At magnetic fields just above (below) the LFSC-IFSC (IFSC-HFSC) transition field, the vortex lattice structure in the IFSC state would be mainly determined by the dominating LFSC (HFSC) order parameter. Then, there may be a phase transition between the two competing vortex structures at magnetic fields in the middle of the IFSC state. If this transition is of first order, there would be domains and domain walls, causing additional vortex pinning.
It should  be noted however that the formation of a nonsingular vortex state in bulk superconductors is still under debate. It has been argued that such a vortex state would be energetically unstable~\cite{Babaev2002,Sigrist1991}.
Since there are only few examples of superconductors with multiple SC states, further studies are needed to understand the vortex dynamics across different SC states in spin-triplet superconductors.

In conclusion, we studied the vortex dynamics in UTe$_2$ by measuring DC-electrical resistivity with different currents along the $a$-axis under magnetic fields along the $b$-axis. We found a region of low critical current deep inside the SC state, which agrees with the recently proposed IFSC state~\cite{Sakai2022arxiv}. The island formation of the low critical current region leads to an unusual maximum of critical current as a function of temperature. The accompanying decrease in the critical current with decreasing temperature cannot be explained by thermal fluctuations but is ascribed to the weakening of the pinning force.  We discuss the possible origins for the weakening of the pinning force, including changes in the SC gap structure and nonsingular vortex states, which may host Majorana fermions~\cite{Read2000,Ivanov2001}.
Further intensive studies are required to understand the anomalous vortex state in spin-triplet superconductors.

We thank J. Goryo, Y. Matsuda, M. Garst, T. Takimoto, Y. Nagai and M. Machida for stimulating discussions. We thank M. Nagai and K. Shirasaki for experimental support. The work was supported by JSPS KAKENHI Grant Numbers, JP20K20905, 20KK0061, JP21H05470, JP22H01176, JP22H00109. This work (A part of high magnetic field experiments) was performed at HFLSM under the IMR-GIMRT program (Proposal Numbers 202012-HMKPB-0012, 202112-HMKPB-0010, 202112-RDKGE-0036 and 202012-RDKGE-0084).

%

\end{document}